\documentstyle[12pt,aas2pp4]{article}

\newcommand {\ASCA}{{\em ASCA}}
\newcommand {\Chandra}{{\em Chandra}}

\slugcomment{Accepted in ApJ}
\lefthead{H. Murakami et al.}
\righthead{A New X-ray Reflection Nebula Sgr C}

\begin{document}

\title{{\ASCA} Discovery of Diffuse 6.4 keV Emission Near the Sgr C
Complex: A New X-ray Reflection Nebula}

\author{Hiroshi~Murakami\altaffilmark{1},
Katsuji~Koyama\altaffilmark{2}, and
Masahiro~Tsujimoto\altaffilmark{1}}
\affil{Department of Physics, Faculty of Science, Kyoto University,
Sakyo-ku, Kyoto 606-8502, Japan; hiro@cr.scphys.kyoto-u.ac.jp,
koyama@cr.scphys.kyoto-u.ac.jp, tsujimot@cr.scphys.kyoto-u.ac.jp}

\and

\author{Yoshitomo~Maeda}
\affil{Department of Astronomy and Astrophysics,
The Pennsylvania State University,
University Park, PA 16802-6305, U.S.A.; maeda@astro.psu.edu}

\and

\author{Masaaki~Sakano}
\affil{Space Utilization Research Program,
National Space Development Agency of Japan, 2-1-1, Sengen, Tsukuba,
Ibaraki, 305-8505, Japan; sakano.masaaki@nasda.go.jp}

\altaffiltext{1}{Research Fellow of the Japan Society for the Promotion 
of Science (JSPS)}
\altaffiltext{2}{CREST, Japan Science and Technology Corporation
(JST), 4-1-8 Honmachi, Kawaguchi, Saitama, 332-0012, Japan}

\begin{abstract}
We present an {\ASCA} discovery of diffuse hard X-ray emission from
the Sgr~C complex with its peak in the vicinity of the molecular cloud
core. The X-ray spectrum is characterized by a strong 6.4-keV line and
large absorption. These properties suggest that Sgr~C is a new {\it
X-ray reflection nebula} which emits fluorescent and scattered X-rays
via irradiation from an external X-ray source. We found no adequately
bright source in the immediate Sgr~C vicinity to fully account for the
fluorescence. The irradiating source may be the Galactic nucleus
Sgr~A$^*$, which was brighter in the past than it is now as is
suggested from observations of the first {\it X-ray reflection nebula}
Sgr~B2.
\end{abstract}


\section{Introduction}
The Galactic Center (GC) region of our Galaxy is a unique
concentration of stars and molecular clouds, in addition to Sgr~A$^*$,
a $2.6 \times 10^6$~M$_\odot$ massive black hole (MBH) at the
gravitational center (for a recent review, see Genzel \&
Eckart~1999\markcite{Genzel99}).

The inner $\simeq 100$~pc region of the Galaxy is a site of a nuclear
starburst and consequently should be populated by a variety of X-ray
sources, such as X-ray binaries, supernova remnants, and young stellar
clusters. {\ASCA} revealed diffuse emissions of $2'$ extent
centered at Sgr~A$^*$ and asymmetrical structure extending $\simeq
1^\circ$ along the Galactic plane (Koyama et
al. 1996\markcite{Koyama96}; Maeda et al. 1998\markcite{Maeda98}). The
spectral characteristics were most remarkable, exhibiting many
K-shell lines from highly ionized atoms, with particularly enhanced
emission from He-like (6.70 keV) and H-like (6.97 keV) irons.

Early {\Chandra} results confirmed the {\ASCA} findings and resolved
numerous X-ray structures: X-ray binaries, diffuse structures,
emission from the supernova remnant Sgr~A-East, emission from
individual stars within the Sgr~A region, and a possible X-ray
counterpart of the
massive black hole Sgr~A$^*$ with a luminosity of only 10$^{33}$
erg~s$^{-1}$ (Baganoff et al. 2000\markcite{Baganoff00}), which is
extremely low compared to typical massive black holes.

Among these many X-ray objects, the most massive molecular cloud
Sgr~B2 in the Galaxy is extraordinary. It exhibits a very peculiar
spectrum with a strong emission line at 6.4 keV, a low energy cutoff
below 4~keV and a pronounced edge-structure at 7.1~keV. The X-ray
image is shifted from the core of the molecular cloud toward the GC by
about 1$\farcm$3 (Murakami et
al. 2000a\markcite{Murakami00a}). Recently {\Chandra} confirmed these
results: the extended structure of the 6.4 keV line image with a convex
shape pointed towards Sgr~A$^*$ and its offset morphology from the
cloud core (Murakami et al. 2000b\markcite{Murakami00b}). A numerical
simulation demonstrated that the offset morphology and the line and
edge dominated spectrum are well reproduced by the reflection of
external X-rays coming from the direction of the GC (Figure 2, Murakami et
al. 2000a\markcite{Murakami00a}; also in Sunyaev et
al. 1998\markcite{Sunyaev98}). Based on this, they proposed Sgr~B2 to be a new
class of X-ray object, an {\it X-ray reflection nebula} (XRN).
However, there is no adequately bright source in the immediate Sgr~B2
vicinity to fully account for the fluorescence. They suspected that
the most likely source is an X-ray outburst from the MBH at Sgr~A$^*$
despite its considerable distance from Sgr~B2 (Koyama et
al. 1996\markcite{Koyama96}; Murakami et al.
2000a\markcite{Murakami00a}). Their model requires an outburst of
$L_{\rm X} \sim 10^{39}$ erg~s$^{-1}$ for at least 10 years and must
have stopped at most 30 years ago to avoid detection by earlier X-ray
astronomical instruments. Such an outburst from a $10^6$~M$_\odot$
MBH, perhaps caused by a surge in accretion rate, is consistent with the
behavior of active galactic nuclei.

The XRN scenario and the putative past X-ray outburst of the Galactic
nucleus Sgr~A$^*$ may indicate that there should be many other XRNs
in the GC region. We therefore examined the archival {\ASCA} data of
the GC region in detail, and found another XRN candidate in the Sgr~C
region, located at the same projected distance as Sgr~B2 from the GC,
but in the opposite direction.

\section{Observation}
The {\ASCA} observations of Sgr~C were made on 1993 October~4. All
four focal plane instruments, two Solid-state Imaging Spectrometers
(SIS0, SIS1) and two Gas Imaging Spectrometers (GIS2, GIS3) were
operated in parallel, providing four independent data sets.
Details of the instruments, the telescopes and the detectors,
are found in Tanaka, Inoue, \& Holt (1994)\markcite{Tanaka94},
Serlemitsos et al. (1995)\markcite{Serlemitsos95}, Burke et
al. (1991)\markcite{Burke91}, Ohashi et al. (1996)\markcite{Ohashi96},
Makishima et al. (1996)\markcite{Makishima96}, and Gotthelf
(1996)\markcite{Gotthelf96}. Each of the GISs was operated in PH mode
with the standard bit-assignment, while the SIS was operated in 4-CCD
bright mode. The data were post-processed to correct for spatial gain
non-linearity. Data taken at geomagnetic cutoff rigidities lower than
6~GV, at elevation angles less than 5$^\circ$ from the earth rim, or
during the passage through the South Atlantic Anomaly were
excluded. After these filterings, the net observing time was 20~ksec.

\section{Analysis and Results}
The key characteristic of XRNs is a strong iron line at 6.4~keV
(Koyama et al. 1996\markcite{Koyama96}; Sunyaev \& Churazov
1998\markcite{Sunyaev98}; Murakami et
al. 2000a\markcite{Murakami00a}), hence we made a narrow energy band
image with a central energy of 6.4~keV and a width of twice the
energy resolution (FWHM): 5.8--7.0~keV for the GIS. Figure~1 shows the
GIS image overlaid with the contour map of the cold cloud density
distribution contour by the radio observation of CS with the radial
velocity of $-120$~km~s$^{-1}$ to $-110$~km~s$^{-1}$ (Figure~3 in
Tsuboi, Handa, \& Ukita 1999\markcite{Tsuboi99}). 
The position accuracy is about 5" for the radio map, and 24" for the
GIS image (Gotthelf et al. 2000\markcite{Gotthelf00}).
The 6.4 keV band image shows an X-ray peak (inner region of the solid
circle) near the molecular cloud core CO~$359.4+0.0$ (Oka et
al. 1998\markcite{Oka98}), which is one of the giant molecular clouds
in the Sgr~C complex, with a mass of $\sim$ 10$^6$~M$_\odot$. We
therefore suspect that the X-ray emission is associated with the
molecular cloud Sgr~C, although a more detailed comparison will require
more accurate positional observations both in X-rays and radio
bands.

We made X-ray spectra using the X-ray photons in a circle of
2$\farcm$7 radius around the X-ray peak (see Figure~1). Since the
Sgr~C region is in the large scale GC plasma
(Koyama et al. 1996\markcite{Koyama96}), the spectrum may be
contaminated by the emission of He- and H-like iron lines (6.70 keV,
and 6.97 keV). In order to properly subtract these highly ionized
iron lines, the background region is selected as a 2$\farcm$7 radius
region with the same Galactic latitude (dotted circle in Figure~1).

Figure~2 shows the background subtracted GIS spectrum. We made a model
spectrum with a power-law continuum and a Gaussian line, folding with
the {\ASCA} response function, and fitted it to the data.  Due to the
limited statistics, we fixed the power-law photon index $\alpha$ to be
2.0 (number of photons $N(E) dE
\propto E^{-\alpha} dE$, where $E$ is the photon energy), the same
value as assumed for Sgr~B2 (Koyama et al. 1996\markcite{Koyama96};
Murakami et al. 2000a\markcite{Murakami00a}). The fit is acceptable,
with the best-fit parameters shown in Table~1. The central energy of
the line appears at 6.28$^{+0.15}_{-0.21}$~keV, in agreement with the
fluorescent line from neutral irons. The line equivalent width and
absorption column density are both very large: $\sim$ 0.8~keV and
1.3$^{+0.3}_{-0.4} \times 10^{23}$~H~cm$^{-2}$, respectively.
Since the Galactic diffuse background may not be uniform, we checked the 
ambiguity of the background selection. We selected three other background
regions with the same Galactic latitude as the source region, and fitted them
with the same model. The best-fit parameters of the line central
energy and the absorption are in the range of statistical errors.
The absorbed flux differs a little larger than the error, however this flux
variation is not serious for the discussion in section 4.

Adding two narrow Gaussian lines at 6.4 and 6.7 keV on a power-law
continuum, we found that the 6.7-keV line from highly ionized iron is
very weak; the upper limit of 9 $\times 10^{-14}$
erg~s$^{-1}$~cm$^{-2}$ is about 1/4 of the best-fit value of the
6.4-keV line. We thus conclude that the emission line from the Sgr~C
cloud is dominated by the fluorescent line from cold irons.

We also made an SIS spectrum from the same region of GIS, and fitted
it with the same model as with the GIS. The best-fit spectral
parameters are shown in Table~1, and are
consistent with those obtained from the GIS spectrum. Due to the
limited statistics, the SIS spectrum does not further constrain the
spectral parameters. We thus discuss the X-ray properties of Sgr~C
using the spectral parameters obtained with GIS only.

\section{Discussion} 
\subsection{Is Sgr~C a New XRN?}
We have found a hard X-ray enhancement near the Sgr~C cloud on the
6.4~keV line map. The spectrum shows a strong fluorescent line at
6.4~keV and the large absorption in low energy band. These properties
are good evidence that Sgr~C is a new XRN (Sunyaev \& Churazov
1998\markcite{Sunyaev98}; Murakami et
al. 2000a\markcite{Murakami00a}).

Murakami et~al. (2000a)\markcite{Murakami00a} analyzed the spectrum of
Sgr~B2, the XRN, and found that the equivalent width of
the 6.4~keV line is 2.9$^{+0.3}_{-0.9}$~keV, and the column density of
hydrogen is 8.3$^{+2.5}_{-2.0} \times 10^{23}$~H~cm$^{-2}$. Those of
Sgr~C are 0.8$^{+0.4}_{-0.5}$~keV and 1.3$^{+0.3}_{-0.4} \times
10^{23}$~H~cm$^{-2}$, respectively.

The absorption of Sgr~C is smaller than that of Sgr~B2 by an order of
magnitude. This is reasonable because the mass of the Sgr~C cloud is
about 1/7 of the Sgr~B2 cloud (Oka et al. 1998\markcite{Oka98}), in
spite of rather similar geometrical size. In fact, using the Sgr~C
cloud size of $\sim$ 28~pc (Oka et~al. 1998\markcite{Oka98}), we can
roughly estimate the column density to be 4 $\times
10^{22}$~H~cm$^{-2}$, which is almost equal to 1.3 $\times
10^{23}$~H~cm$^{-2}$, after taking account of the interstellar
absorption to the Sgr~C cloud of $\sim 10^{23}$~H~cm$^{-2}$ (Sakano et
al. 1998\markcite{Sakano98}).

The absorption column of 4 $\times 10^{22}$ H~cm$^{-2}$ is in the optically
thin range near 6.4~keV and above 7.1~keV energy, hence, unlike
Sgr~B2, the 6.4~keV emission region should overlap the molecular
core. This is in agreement with the observation, although the
observational results still have significant uncertainty.

\subsection{Irradiating Source of XRNs}
In the optically thin case, the X-ray intensity of an XRN
is simply proportional to the amount of the scattering matter
and the flux of the irradiating X-rays. The absorption-corrected
luminosity of the 6.4~keV line from Sgr~C is 4 $\times 10^{33}$
erg~s$^{-1}$ within a 2$\farcm$7-radius circular region. 
The required luminosity of the X-ray source
irradiating Sgr~C would be about 3 $\times 10^{39}$ ($d_{\rm SgrC}$/100
pc)$^2$ erg~s$^{-1}$, where $d_{\rm SgrC}$ is the distance from the
irradiating X-ray source to Sgr~C.

The brightest nearby X-ray source is 1E~1740.7$-$2942, which lies only
$\sim$ 0.4 degrees from Sgr~C. During the present observation, its
luminosity was $\sim 3 \times 10^{36}$~erg~s$^{-1}$ in the 2--10~keV
band (Sakano et al. 1999\markcite{Sakano99}). This luminosity,
however, is two orders of magnitude less than that required to account
for the fluorescent X-rays from Sgr~C. Even if we accumulate the X-ray
fluxes from all cataloged bright X-ray sources near the GC (Sakano
2000\markcite{Sakano00}), we can explain only 2 \% of the reflected
luminosity of Sgr~C. This fact strengthens the scenario that the
X-rays from Sgr~C, as well as those from Sgr~B2, is due to the
reflection from an irradiating source which was very bright in the
past but is presently dim. The X-ray fluxes of Sgr~B2 and Sgr~C can be
consistently explained by one irradiating source, if it is located at
almost the same distance from both of the XRNs. This position falls
near the Galactic nucleus Sgr~A$^*$, and the X-ray luminosity should
be as luminous as 3 $\times 10^{39}$~erg~s$^{-1}$ about 300 years ago,
the light travel time from Sgr~A$^*$ to Sgr~B2 and Sgr~C.

\section{Summary}
1. We have discovered diffuse emission in the 6.4 keV band from the
Sgr~C region based on the {\ASCA} observations.

2. We found that the X-ray spectrum of Sgr~C shows a strong emission
line at 6.4 keV from cold iron with an equivalent width of about 0.8
keV, and a large absorption of $1.3 \times 10^{23}$ H cm$^{-2}$.

3. The line dominated spectrum and the image correlated with a
molecular cloud are similar to those of
the prototypical X-ray reflection nebula (XRN) Sgr~B2, and thus Sgr~C
can be regarded as a new XRN.

4. There is no adequately bright source in the immediate Sgr~C vicinity to
fully account for the observed fluorescence.

5. Existence of the two XRNs, Sgr~B2 and Sgr~C, can be fully explained if we
consider the AGN activity some hundreds of years ago at the GC ($L_{\rm X}
\sim 3 \times 10^{39}$ erg s$^{-1}$).

\vspace*{1em}

\acknowledgements
The authors express their thanks to all the members of the {\ASCA}
team. H.M and M.T. are financially supported by the Japan Society for
the Promotion of Science. Thanks are also due to Patrick Durrell for a
critical reading of the manuscript.

\newpage
\onecolumn

\begin{figure}
\epsscale{0.8}
\plotone{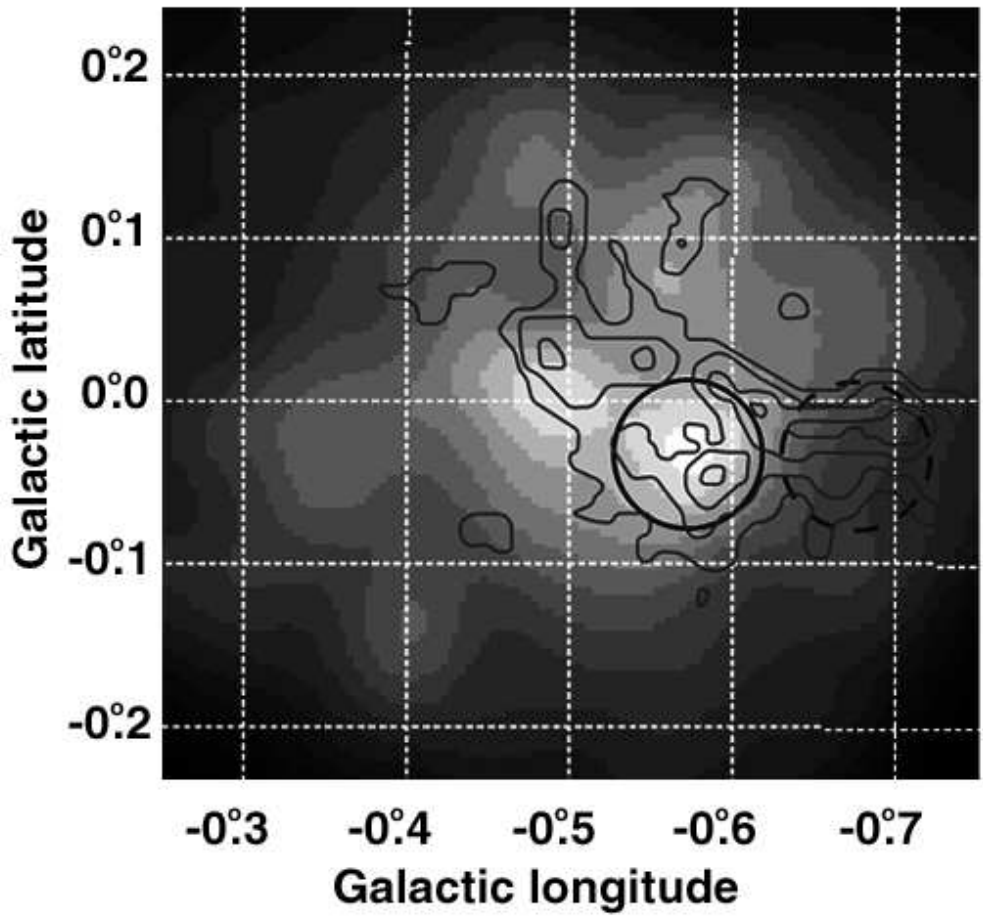}
\caption[f1.ps]{The 5.8--7.0~keV band image around the Sgr~C
cloud obtained with the GIS, laid over the CS line contours
(Figure 3 in Tsuboi, Handa, \& Ukita 1999).
The solid and the dotted circle show the source and the
background regions, respectively.}
\end{figure}

\begin{figure}
\plotone{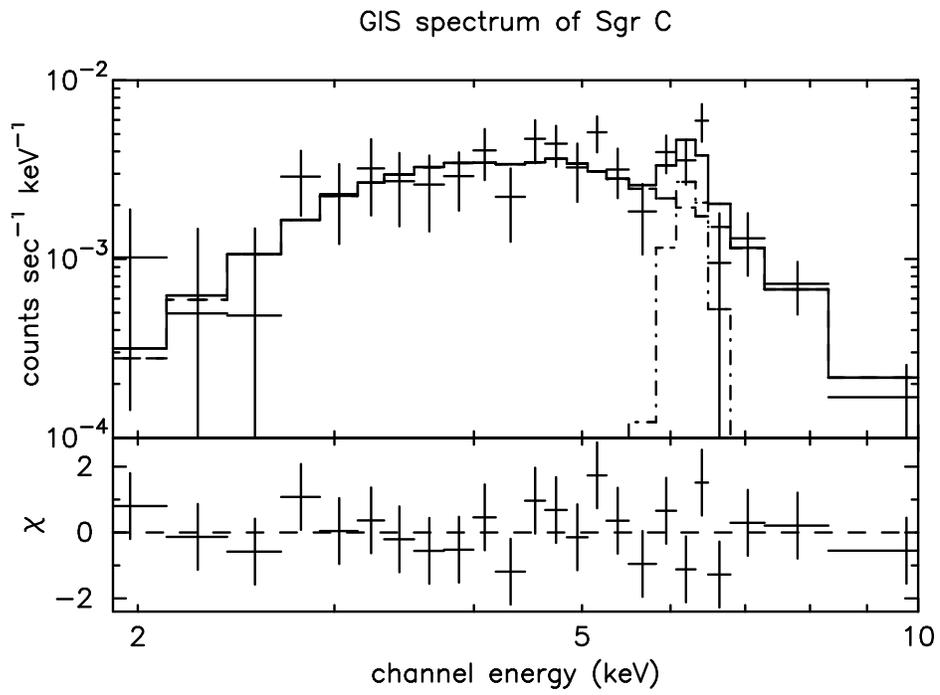}
%\plotfiddle{f2.ps}{18cm}{0}{100}{100}{0}{0}
\caption[f2.ps]{The GIS (GIS2 + 3) spectrum of Sgr~C. The solid
line shows the best-fit model shown in Table~1.}
\end{figure}

\begin{deluxetable}{llccc}
\tablecaption{Fitting Results of Sgr C to a Phenomenological Spectral Model}
\tablehead{
\colhead{Model Components}   &\colhead{Parameters}   &\colhead{Unit} 
		&\colhead{GIS}		&\colhead{SIS}}
\startdata
Absorption	& $N_{\rm H}$\tablenotemark{a} & (10$^{22}$~H~cm$^{-2}$)
		& 12.6$^{+3.5}_{-3.3}$	& 9.7$^{+3.9}_{-2.5}$\nl
Continuum	& Photon Index &     
		& 2.0 (fixed)		& 2.0 (fixed) \nl
		& Flux\tablenotemark{b} (2--10 keV) & (10$^{-4}$~ph~s$^{-1}$~cm$^{-2}$)
		& 3.0$^{+0.9}_{-0.6}$	& 2.0$^{+0.7}_{-0.5}$\nl
Fe Line 	& Center Energy	& (keV)	
		& 6.28$^{+0.15}_{-0.21}$& 6.38$^{+0.10}_{-0.31}$\nl
		& Flux\tablenotemark{b} & (10$^{-5}$~ph~s$^{-1}$~cm$^{-2}$)
		& 3.5$^{+1.4}_{-2.2}$	& 2.7$^{+2.3}_{-1.5}$\nl
		& Equivalent Width	& (keV) 
		& 0.8$^{+0.4}_{-0.5}$	& 1.1$^{+1.1}_{-0.7}$\nl
\tableline
Total Luminosity& $L_{4-10 {\rm keV}}$	& (10$^{34}$~erg s$^{-1}$)
		& 5.0 			& 3.0 \nl 
\tableline
Reduced $\chi^2$ (d.o.f.)& &		
		& 0.80 (20)    		& 0.63 (14) \nl
\enddata
\tablecomments{The errors are at 90\% confidence level.}
\tablenotetext{a}{The equivalent hydrogen column density for the solar 
abundances.}
\tablenotetext{b}{The fluxes are not corrected for absorption.}
\end{deluxetable}

\end{document}